\documentclass[runningheads]{llncs}
\usepackage[table]{xcolor}
\usepackage{graphicx}
\usepackage[hidelinks]{hyperref}
\usepackage{amsmath,bm}
\usepackage{dsfont}
\usepackage{siunitx}
\usepackage{algorithm2e}

\PassOptionsToPackage{hyphens}{url}\usepackage{hyperref}
\usepackage{amssymb}
\usepackage{todonotes}
\usepackage{float}
\usepackage{ circuitikz }
\usepackage[font=small,skip=2pt]{caption}
\usepackage{booktabs}

\usepackage{listings}

\definecolor{darkred}{RGB}{193, 39, 45}
\definecolor{indigo}{RGB}{0, 0, 167}
\definecolor{teal}{RGB}{0, 129, 118}
\definecolor{yellow}{RGB}{238, 204, 22}
\definecolor{lightgray}{RGB}{179, 179, 179}

\usepackage{mathtools}

\begin{document}
\title{A HPX Communication Benchmark: Distributed FFT using Collectives}
\titlerunning{A HPX Communication Benchmark}
% If the paper title is too long for the running head, you can set
% an abbreviated paper title here
%
% If the paper title is too long for the running head, you can set
% an abbreviated paper title here
\author{Alexander Strack\inst{1}\orcidID{0000-0002-9939-9044}  
\and Dirk Pflüger\inst{1} \orcidID{0000-0002-4360-0212}}
\authorrunning{A. Strack \and D. Pflüger}
% First names are abbreviated in the running head.
% If there are more than two authors, 'et al.' is used.
%
\institute{Institute of Parallel and Distributed Systems, University of Stuttgart,\\ 70569 Stuttgart, Germany\\ \email{\{alexander.strack, dirk.pflueger\}@ipvs.uni-stuttgart.de}
}
\maketitle              % typeset the header of the contribution
\begin{abstract}
Due to increasing core counts in modern processors, several task-based runtimes emerged,
including the C\texttt{++} Standard Library for Concurrency and Parallelism (HPX). 
Although the asynchronous many-task runtime HPX allows implicit communication via an Active Global Address Space, it also supports explicit collective operations. Collectives are an efficient way to realize complex communication patterns. 

In this work, we benchmark the TCP, MPI, and LCI communication backends of HPX, which are called parcelports in HPX terms. We use a distributed multi-dimensional FFT application relying on collectives. Furthermore, we compare the performance of the HPX \emph{all-to-all} and \emph{scatter} collectives against an FFTW3 reference based on \emph{MPI+X} on a 16-node cluster.

Of the three parcelports, LCI performed best for both \emph{scatter} and \emph{all-to-all} collectives. 
Furthermore, the LCI parcelport was up to factor 3 faster than the \emph{MPI+X} reference.
Our results highlight the potential of message abstractions and the parcelports of HPX.

\end{abstract}

\setcounter{footnote}{0} 
\section{Introduction}\label{sec:introduction}
The use of collective operations to simplify the communication between distributed memory environments is common in high-performance computing. Implementations of the Message Passing Interface (MPI), e.g., OpenMPI \cite{Graham2006}, provide several highly optimized collectives.
Recently several asynchronous many-task runtimes emerged, as alternatives to the standard parallelization model based on MPI+X. See  \cite{Thoman2018} for a detailed comparison of existing asynchronous many-task runtimes. In this work, we focus on HPX, an actively developed C\texttt{++} Standard Library for Concurrency and Parallelism \cite{hpx}. While HPX realized asynchronous parallelism via futurized tasks, its Active Global Address Space (AGAS) provides distributed functionalities. In contrast to messages, HPX uses an abstraction called \emph{parcel}. These \emph{parcels} are transferred over a so-called parcelport. HPX supports multiple parcelports, allowing to switch the communication backend via command line arguments.

We use collectives to implement a distributed Fast Fourier Transform (FFT) and benchmark the parcelports of HPX against the \emph{MPI+X} parallelization of the popular FFTW3 library \cite{Frigo2005}.

\section{Application}\label{sec:application}
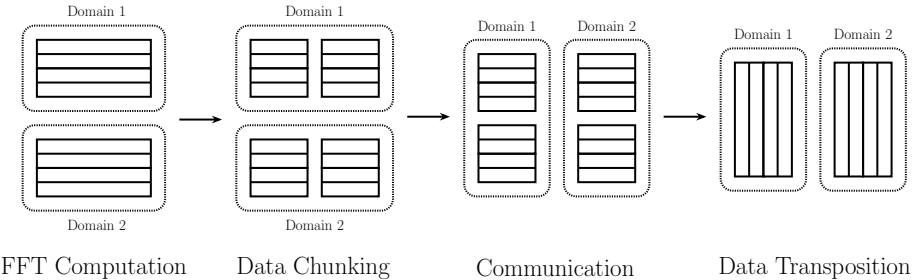
\begin{figure}
    \centering
    \resizebox{1\textwidth}{!}{%
        \begin{circuitikz}
            \tikzstyle{every node}=[font=\LARGE]
            \node [font=\Huge, scale=2.5] at (2.5,-2) {FFT Computation};
            \node [font=\Huge, scale=2.5] at (21.75,-2) {Data Chunking};
            \node [font=\Huge, scale=2.5] at (43,-2) {Communication};
            \node [font=\Huge, scale=2.5] at (65.75,-2) {Data Transposition};
            % First
            \node [font=\Huge, scale=1.5] at (2.75,20.5) {Domain 1};
            \draw [, line width=5pt , rounded corners = 30.0, dashed] (-3.75,19.25) rectangle  (8.75,11.75);
            \node [font=\Huge, scale=1.5] at (2.75,1.75) {Domain 2};
            \draw [, line width=5pt , rounded corners = 30.0, dashed] (-3.75,10.5) rectangle  (8.75,3);
            % Second
            \node [font=\Huge, scale=1.5] at (21.875,20.5) {Domain 1};
            \draw [, line width=5pt , rounded corners = 30.0, dashed] (15,19.25) rectangle  (28.75,11.75);
            \node [font=\Huge, scale=1.5] at (21.875,1.75) {Domain 2};
            \draw [, line width=5pt , rounded corners = 30.0, dashed] (15,10.5) rectangle  (28.75,3);
            % Third
            \node [font=\Huge, scale=1.5] at (38.75,19.25) {Domain 1};
            \draw [, line width=5pt , rounded corners = 30.0, dashed] (35,18) rectangle  (42.5,4.25);
            \node [font=\Huge, scale=1.5] at (47.5,19.25) {Domain 2};
            \draw [, line width=5pt , rounded corners = 30.0, dashed] (43.75,18) rectangle  (51.25,4.25);
            % Fourth
            \node [font=\Huge, scale=1.5] at (61.25,18.5) {Domain 1};    
            \draw [, line width=5pt , rounded corners = 30.0, dashed] (57.5,17.25) rectangle  (65,4.75);
            \node [font=\Huge, scale=1.5] at (70,18.5) {Domain 2}; 
            \draw [, line width=5pt , rounded corners = 30.0, dashed] (66.25,17.25) rectangle  (73.75,4.75);
            % Arrows
            \draw [line width=5pt, ->, >=Stealth] (10,11) -- (13.75,11);
            \draw [line width=5pt, ->, >=Stealth] (30,11.25) -- (33.75,11.25);
            \draw [line width=5pt, ->, >=Stealth] (52.5,11.25) -- (56.25,11.25);
            % Data
            \draw [, line width=5pt ] (-2.5,18) rectangle (7.5,15.5);
            \draw [, line width=5pt ] (-2.5,15.5) rectangle (7.5,13);
            \draw [, line width=5pt ] (-2.5,9.25) rectangle (7.5,6.75);
            \draw [, line width=5pt ] (-2.5,6.75) rectangle (7.5,4.25);
            \draw [, line width=5pt ] (16.25,15.5) rectangle (21.25,13);
            \draw [, line width=5pt ] (22.5,15.5) rectangle (27.5,13);
            \draw [, line width=5pt ] (22.5,18) rectangle (27.5,15.5);
            \draw [, line width=5pt ] (16.25,18) rectangle (21.25,15.5);
            \draw [, line width=5pt ] (16.25,6.75) rectangle (21.25,4.25);
            \draw [, line width=5pt ] (22.5,6.75) rectangle (27.5,4.25);
            \draw [, line width=5pt ] (22.5,9.25) rectangle (27.5,6.75);
            \draw [, line width=5pt ] (16.25,9.25) rectangle (21.25,6.75);
            \draw [, line width=5pt ] (36.25,16.75) rectangle (41.25,14.25);
            \draw [, line width=5pt ] (36.25,14.25) rectangle (41.25,11.75);
            \draw [, line width=5pt ] (36.25,10.5) rectangle (41.25,8);
            \draw [, line width=5pt ] (36.25,8) rectangle (41.25,5.5);
            \draw [, line width=5pt ] (45,16.75) rectangle (50,14.25);
            \draw [, line width=5pt ] (45,14.25) rectangle (50,11.75);
            \draw [, line width=5pt ] (45,10.5) rectangle (50,8);
            \draw [, line width=5pt ] (45,8) rectangle (50,5.5);
            \draw [, line width=5pt ] (58.75,16) rectangle (61.25,6);
            \draw [, line width=5pt ] (61.25,16) rectangle (63.75,6);
            \draw [, line width=5pt ] (67.5,16) rectangle (70,6);
            \draw [, line width=5pt ] (70,16) rectangle (72.5,6);
            \draw [line width=5pt, short] (-2.5,16.75) -- (7.5,16.75);
            \draw [line width=5pt, short] (-2.5,14.25) -- (7.5,14.25);
            \draw [line width=5pt, short] (-2.5,8) -- (7.5,8);
            \draw [line width=5pt, short] (-2.5,5.5) -- (7.5,5.5);
            \draw [line width=5pt, short] (16.25,16.75) -- (21.25,16.75);
            \draw [line width=5pt, short] (16.25,14.25) -- (21.25,14.25);
            \draw [line width=5pt, short] (22.5,16.75) -- (27.5,16.75);
            \draw [line width=5pt, short] (22.5,14.25) -- (27.5,14.25);
            \draw [line width=5pt, short] (16.25,8) -- (21.25,8);
            \draw [line width=5pt, short] (16.25,5.5) -- (21.25,5.5);
            \draw [line width=5pt, short] (22.5,8) -- (27.5,8);
            \draw [line width=5pt, short] (22.5,5.5) -- (27.5,5.5);
            \draw [line width=5pt, short] (36.25,15.5) -- (41.25,15.5);
            \draw [line width=5pt, short] (36.25,13) -- (41.25,13);
            \draw [line width=5pt, short] (36.25,9.25) -- (41.25,9.25);
            \draw [line width=5pt, short] (36.25,6.75) -- (41.25,6.75);
            \draw [line width=5pt, short] (45,15.5) -- (50,15.5);
            \draw [line width=5pt, short] (45,13) -- (50,13);
            \draw [line width=5pt, short] (45,9.25) -- (50,9.25);
            \draw [line width=5pt, short] (45,6.75) -- (50,6.75);
            \draw [line width=5pt, short] (60,16) -- (60,6);
            \draw [line width=5pt, short] (62.5,16) -- (62.5,6);
            \draw [line width=5pt, short] (68.75,16) -- (68.75,6);
            \draw [line width=5pt, short] (71.25,16) -- (71.25,6);
        \end{circuitikz}
    }%
    \caption{Four steps that need to be executed in sequence for each dimension of a two-dimensional FFT on two separated memory domains.} 
    \label{fig:steps}
\end{figure}

The FFT is one of the most common algorithms in modern computing.
The one-dimensional transform can be easily expanded into multiple dimensions by dimension-wise FFT computations. 
In order to compute a distributed FFT in multiple dimensions, four steps are necessary (see Figure \ref{fig:steps}).
Hereby, the communication step is the most expensive. Assuming $N$ participating nodes, the communication step requires the transfer of $(1-\frac{1}{N})$ chunks of the equally chunked local data. More specifically, the $i$-th chunk needs to be transferred to the $i$-th participating node. 
As point-to-point communication becomes complicated quickly, this communication pattern is realized with collectives in practice.

\section{Methods}\label{sec:methods}
While there exist various collectives, we focus on those relevant to the FFT communication pattern. The easiest way to implement this pattern is the \emph{all-to-all} collective. While \emph{all-to-all} exactly matches the FFT communication pattern, it has one major drawback: The \emph{all-to-all} collective is synchronized. However, the arriving data chunks can be transposed as soon as they are received. Thus, we propose to replace the \emph{all-to-all} collective with a combination of $N$-\emph{scatter} collectives. 
This allows us to hide some of the computation behind the long communication time.

At the time of writing, HPX ships with three parcelports:
\begin{itemize}
    \item TCP Parcelport: Requires no external dependencies.
    \item MPI Parcelport: Requires external MPI installation.
    \item LCI Parcelport: Can be fetched alongside HPX or requires external LCI installation.
\end{itemize}
Both TCP \cite{Cerf1974} and MPI \cite{Mpi2012} parcelports were provided by Heller \cite{Heller2019}. The TCP parcelport is mainly intended to act as a fallback in case something goes wrong. Yan et. al. recently added support for a new parcelport to HPX based on the \emph{Lightweight Communication Interface} (LCI) \cite{Yan23}.

\section{Results}\label{sec:results}
\begin{figure}[h]
\centering
    %X\dotfill X\newline
    \begin{minipage}[c]{0.34\hsize}\centering
        \begin{tabular}{lll}
        \toprule
        Cluster            & buran                  \\ \midrule
        %\rowcolor{lightgray}
        Nodes           & 16                             \\ %\hline
        \rowcolor{lightgray!50}
        Connection      & InfiniBand HDR   \\ %\hline
        %\rowcolor{lightgray}
        Speed           & 200Gb/s                   \\ %\hline
        \rowcolor{lightgray!50}
        Sockets  & 2                             \\ %\hline
        %\rowcolor{lightgray}
        CPU  & AMD EPYC 7352 \\ %\hline
        \rowcolor{lightgray!50}
        Cores           & 24                            \\ %\hline
        %\rowcolor{lightgray}
        Clock rate       & 2.3GHz                    \\ %\hline
        \rowcolor{lightgray!50}
        L3 Cache        & 128MB                     \\ %\hline
        %\rowcolor{lightgray}
        RAM             & 256GB                     \\ \bottomrule
        \end{tabular}\newline\newline
    \caption{Hardware specification of benchmark cluster}
    \label{fig:specs}
    \end{minipage}
    \hfill
    \begin{minipage}[c]{0.60\hsize}\centering
        \centering
        %\raggedleft
        \includegraphics[width=\linewidth]{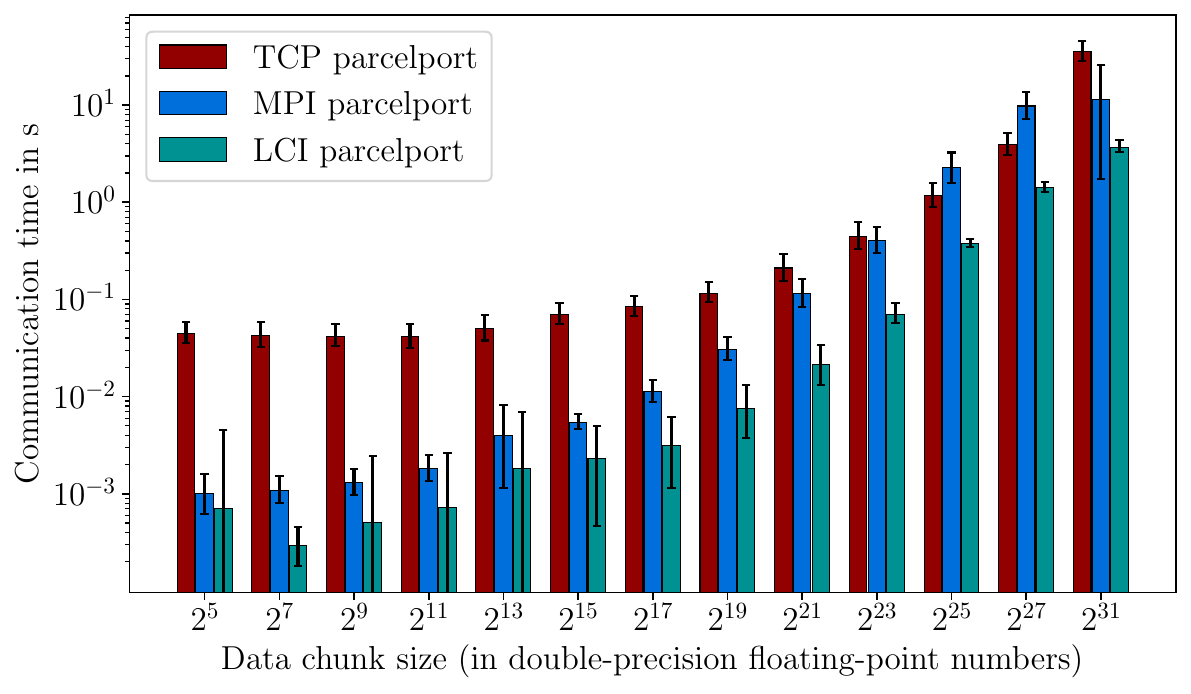}
        \caption{Chunk size scaling on two nodes.}
        \label{fig:buran_message}
    \end{minipage}%
\end{figure}

The results in this section were obtained on a 16-node cluster. For more detailed hardware specifications, see Figure \ref{fig:specs}. All runtimes are averaged over 50 runs and are visualized with $95\%$ confidence bars.
\begin{figure}[bp]
    \centering
    \begin{minipage}[t]{.47\textwidth}
        \centering
        %\raggedright
        \includegraphics[width=\linewidth]{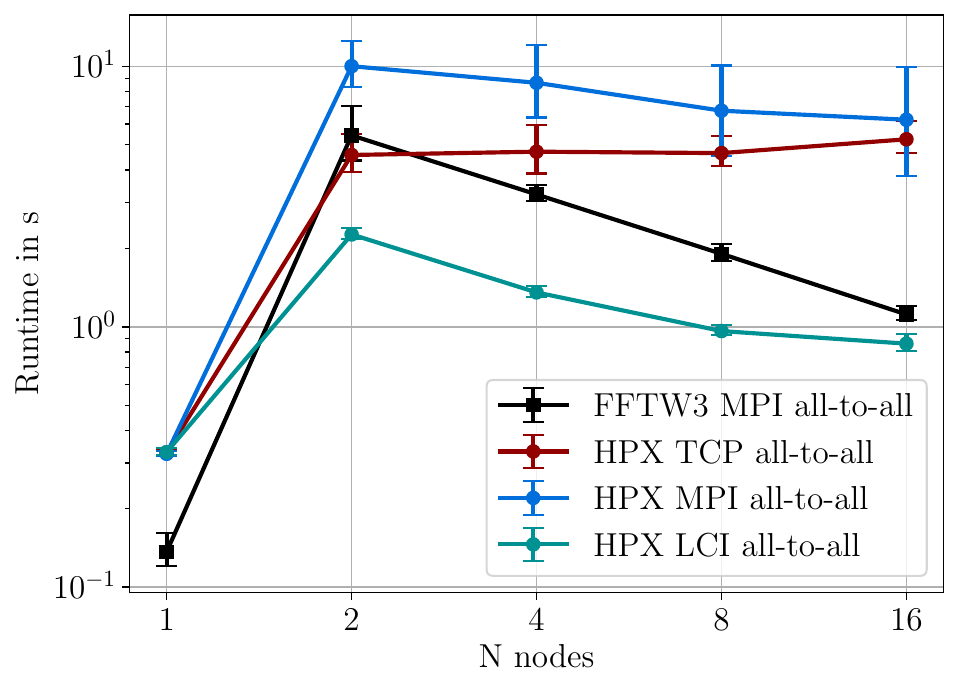}
        \caption{Strong scaling on up to 16 nodes for HPX \emph{all-to-all} collective.}
        \label{fig:buran_all_to_all}
    \end{minipage}\hspace{.05\textwidth}
    \begin{minipage}[t]{.47\textwidth}
        \centering
        %\raggedleft
        \includegraphics[width=\linewidth]{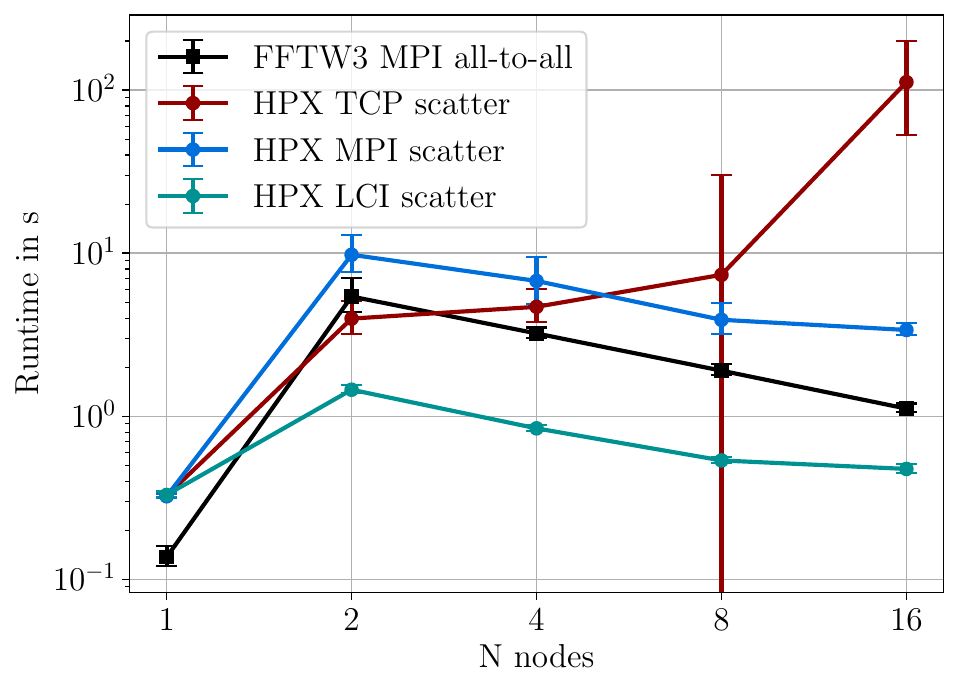}
        \caption{Strong scaling on up to 16 nodes for HPX \emph{scatter} collective.}
        \label{fig:buran_scatter}
    \end{minipage}
\end{figure}

In our chunk size benchmark (see Figure \ref{fig:buran_message}), we use the \emph{scatter} collective to simulate two separate one-way communication channels between two nodes. The LCI parcelport outperforms the other parcelports across all chunk sizes. Especially the TCP parcelport has a big overhead for small data chunks. 

In our strong scaling benchmarks, we compare the parcelports for both \emph{all-to-all} (see Figure \ref{fig:buran_all_to_all}) and \emph{scatter} (see Figure \ref{fig:buran_scatter}) collectives against an FFTW3 implementation parallelized with \emph{MPI+pthreads}. We choose a two-dimensional FFT of size $2^{14} \times 2^{14}$. 
Generally, the \emph{scatter} based approach is faster than one \emph{all-to-all} collective. Interestingly, the TCP parcelport is on the benchmark cluster faster than the MPI parcelport for the used problem size. However, the runtimes of the TCP parcelport skyrocket for the \emph{scatter} approach. Again the LCI parcelport dominates the performance and is even able to beat the FFTW3 reference up to a factor of 3.

\section{Conclusion and outlook}\label{sec:conclusion}
In the scope of this work, we used distributed FFT to evaluate the performance of the existing HPX parcelports for different chunk sizes. Furthermore, we compared the performance of HPX collectives against the popular FFTW3 based on MPI collectives. 
The $N$-\emph{scatter} approach induces less overhead in HPX and allows for more asynchronous parallelism. Thus, it can help to accelerate similar applications. 
Even though the HPX collectives are not optimized to rival their MPI equivalents in direct comparison, the LCI parcelport accelerated the communication to such an extent, that our HPX implementation beat FFTW3 up to a factor of 3.

These results highlight the flexibility and the potential of the parcelport concept used in HPX. We plan to provide more parcelports for HPX in order to add even more flexibility for users.

\section*{Supplementary materials}
The source code and benchmark scripts are available at {DaRUS}\footnote{\url{https://doi.org/10.18419/darus-4520}}. The respective software and compiler versions are stated in the \lstinline[language=bash]{README.md} of the source code. 

% ---- Bibliography ----
\bibliographystyle{splncs04}
\bibliography{bibliography}
\end{document}